\documentclass[twocolumn,prl,showpacs]{revtex4}
\usepackage{graphicx}
\usepackage{hyperref}
\usepackage{float,calc,xspace,units}
\usepackage{amsmath,amsfonts,amssymb}
\usepackage[small,bf]{caption}

\catcode`\ä = \active \catcode`\ö = \active \catcode`\ü = \active
\catcode`\Ä = \active \catcode`\Ö = \active \catcode`\Ü = \active
\catcode`\ß = \active
\defä{\"a}
\defö{\"o}
\defü{\"u}
\defÄ{\"A}
\defÖ{\"O}
\defÜ{\"U}
\defß{\ss}
\def\fm#1{\ifmmode #1 \else $#1$\fi}

\def\vecr{\fm{\vec{r}}\xspace}
\def\kB{\fm{k_\mathrm{B}}\xspace}
\def\wr{\fm{\omega_r}\xspace}
\def\wx{\fm{\omega_x}\xspace}
\def\wy{\fm{\omega_y}\xspace}
\def\wz{\fm{\omega_z}\xspace}
\def\taurel{\fm{\tau_\mathrm{rel}}\xspace}
\def\Garel{\fm{\Gamma_\mathrm{rel}}\xspace}
\def\Gacoll{\fm{\Gamma_\mathrm{coll}}\xspace}

\def\DTts{\fm{\Delta T(t^*)}\xspace}
\def\vrth{\fm{\EW{v_r}_\mathrm{th}}\xspace}
\def\uK{\fm{\mu\mathrm{K}}\xspace}
\def\sigmaeff{\fm{\sigma_\mathrm{eff}}\xspace}
\def\EW#1{{\ifmmode <\!#1\!>\else $<\!#1\!>$\fi}}
\def\Crzf{\fm{{}^{52}\mathrm{Cr}}\xspace}
\def\Crf{\fm{{}^{50}\mathrm{Cr}}\xspace}
\def\af{\fm{a(\Crf)}\xspace}
\def\azf{\fm{a(\Crzf)}\xspace}

\begin{document}

\title{Determination of the $s$-wave Scattering Length of Chromium}
\author{P.~O. Schmidt}\email{p.schmidt@physik.uni-stuttgart.de}
\author{S. Hensler}
\author{J. Werner}
\author{A. Griesmaier}
\author{A. Görlitz}
\author{T. Pfau}
\affiliation{5. Physikalisches Institut, Universität Stuttgart,
70550 Stuttgart, Germany}
\author{A. Simoni}
\affiliation{Dipartimento di Chimica, Universit\'{a} di Perugia,
06123 Perugia, Italy}
\date{\today}

\begin{abstract}
We have measured the deca-triplet $s$-wave scattering length of
the bosonic chromium isotopes \Crzf and \Crf. From the time
constants for cross-dimensional thermalization in atomic samples
we have determined the magnitudes $|\azf|=\unit[(170 \pm
39)]{a_0}$ and $|\af|=\unit[(40 \pm 15)]{a_0}$, where $a_0 =
\unit[0.053]{nm}$. By measuring the rethermalization rate of \Crzf
over a wide temperature range and comparing the temperature
dependence with the effective-range theory and single-channel
calculations, we have obtained strong evidence that the sign of
\azf is positive. Rescaling our \Crzf model potential to \Crf
strongly suggests that \af is positive, too.
\end{abstract}
 \vskip1pc

\maketitle

With the development of laser cooling and trapping techniques,
atomic collisional properties in the ultracold regime have become
directly accessible. Today these properties play a crucial role
for the realization of Bose-Einstein condensates (BECs) and
quantum degenerate samples of fermions by evaporative cooling
\cite{Anglin:2002}. In the ultracold regime, elastic collisions
between neutral atoms are dominated by $s$-wave scattering which
can be characterized by a single parameter, the $s$-wave
scattering length. Pure and stable BECs could so far only be
realized if the scattering length is positive and has a value in
the range from 50 to $\unit[300]{a_0}$. Exceptions are $^7$Li
\cite{Bradley1995a} with a negative scattering length which forms
only condensates with limited number and hydrogen
\cite{Fried1998a} with a very small scattering length where
condensate fractions of no more than \unit[5]{\%} have been
reached. In $^{85}$Rb \cite{Cornish2000a} and $^{133}$Cs
\cite{Weber:2003a}, Feshbach resonances have been used to tune the
scattering length into the right range for achieving BEC.
Therefore, a careful measurement of the scattering length is a
first requirement before a strategy for the condensation of a new
atomic species can be devised.

However, elastic collisions are not only essential for evaporative
cooling of an atomic gas but they also determine the interaction
in a quantum degenerate gas. The presence of this interaction is
also responsible for many of the fascinating features of BECs like
superfluid behavior or the existence of phonon-like excitations
\cite{Anglin:2002}. In the case of chromium, the magnitude of the
scattering length will also determine how significant the effects
of the anisotropic dipole-dipole interaction will be
\cite{Goral2000a}.

Up to now, only the $s$-wave scattering lengths for alkali atoms,
hydrogen and metastable helium have been measured. The most
precise methods are probably photo-association spectroscopy
\citep{Weiner:1999a,Heinzen1999a} and Raman spectroscopy
\cite{Samuelis:2001a} between vibrational levels of the electronic
molecular ground state. Another possibility is Feshbach resonance
spectroscopy \citep{Chin:2000a,Leo:2000a,Marte:2002a} where the
position and magnetic field dependence of several Feshbach
resonances is determined.

For chromium, no molecular spectroscopy data are available for the
deca-triplet state which -- neglecting the spin-spin interaction
-- corresponds to the collisional channel for elastic collisions
in samples prepared in spin-stretched (i.e. $|m_J| = J$) states.
Due to the complicated electronic structure with six unpaired
outer-shell electrons, accurate \textit{ab initio} calculations of
the molecular potentials are also not available. Fortunately
near-threshold collisions are not sensitive to the details of the
short-range interaction potential. They are to a large extent
determined by the $s$-wave scattering length $a$ and by the
long-range dispersion interaction, characterized by the leading
van der Waals coefficient $C_6$.

In this paper, we report on a measurement of the $s$-wave
scattering length of bosonic chromium atoms, spin polarized in the
weak-field seeking $J=m_J=3$ state, from cross-dimensional
relaxation in a magnetic trap. By recording the relaxation of an
anisotropic temperature distribution in an atomic cloud towards
equilibrium, we obtain a relaxation time constant which is
proportional to the effective scattering cross-section through the
atomic density. This method has been first used by Monroe
\textit{et al.}
 \cite{Monroe1993a} for $^{133}$Cs and has since been employed to
determine the ultra-cold scattering properties of many other
alkaline metals. In all of these experiments, only the scattering
cross section, i.e. the magnitude of the scattering length could
be determined, but not its sign. As shown by Ferrari \textit{et
al.} \cite{Ferrari:2002a}, also the sign of the scattering length
can in some instances be determined by a careful measurement of
the temperature dependence of the scattering cross section.

We have determined values of $|\azf|=\unit[(170 \pm 39)]{a_0}$ and
$|\af|=\unit[(40 \pm 15)]{a_0}$. A careful analysis of the
temperature-dependence of the cross section for \Crzf in a range
from 5 to $\unit[500]{\mu K}$ allowed us to compare the
experimental data to the effective-range theory. From this
comparison, we have obtained strong evidence that the scattering
length of \Crzf is positive. Due to its much lower natural
abundance, we could only measure the cross section for \Crf over a
smaller temperature range. Rescaling a \Crzf model potential to
\Crf strongly suggests a positive scattering length also for \Crf.

 The experimental setup is
described in detail in \cite{Schmidt:2003a,Schmidt:2003c} and is
only outlined here. We prepare magnetically trapped,
spin-polarized clouds of chromium atoms by using our CLIP trap
loading procedure \cite{Schmidt:2003a}. After loading, the atoms
are Doppler cooled in the magnetic trap \cite{Schmidt:2003c}.
Further cooling is achieved by forced rf-evaporation in a magnetic
trap with an offset field of \unit[4]{G} and trap frequencies of
$\wr=\wx=\wy=2\pi\times\unit[124]{Hz}$ and
$\wz=2\pi\times\unit[72.6]{Hz}$, where the temperature can be
adjusted by varying the end frequency of the rf ramp.
Subsequently, the magnetic offset field is ramped within
\unit[25]{ms} from 4 to \unit[1.75]{G} which increases $\wr$ to
$2\pi\times\unit[207]{Hz}$. Since the time scale for the
modification of the trap is much faster than the mean time between
collisions, this creates an anisotropic temperature distribution
$\Delta T(t)$. To observe rethermalization, absorption images in
the trap are taken after variable wait times between \unit[10]{ms}
and \unit[2]{s}. The radial and axial temperatures can then be
inferred from the corresponding sizes of the cloud and the known
trap frequencies. The number of atoms and thus the density is
calibrated by taking additional absorption images after release of
the atoms from the trap.

If no additional heating is present, the anisotropic distribution
relaxes to the new steady state temperature
$T^f=T^f_r=T^f_z=(2T^i_r+T^i_z)/3$ via elastic collisions. In the
real experimental situation, inelastic processes dominated by
dipolar relaxation rates cause heating and atom loss, with a
significant decrease of the gas density during relaxation. If the
typical timescale for atom loss is longer than the relaxation
time, the density decrease can be accounted for by employing a
time rescaling method \cite{Hopkins:2000a}. In our relaxation
measurements, the lifetime of the atomic sample exceeded the
relaxation time constant by at least a factor of 10.

\begin{figure}
\includegraphics[width=0.9\columnwidth]{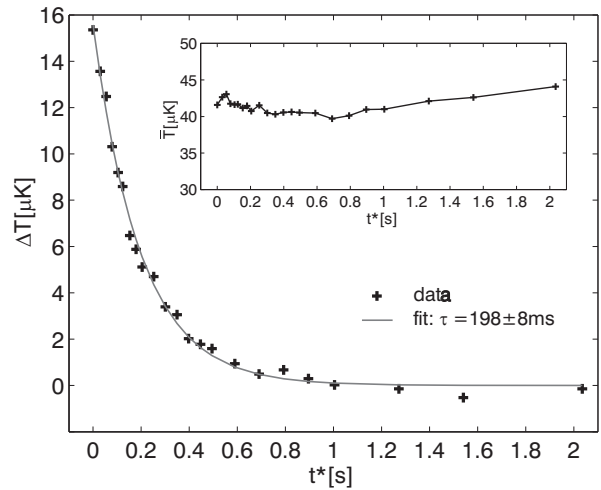}
\caption{Thermal relaxation versus rescaled time $t^*$ of a cloud
of \Crzf atoms in a magnetic trap. The initial anisotropic
temperature distribution is created by a sudden change of the
radial trap frequency from $\wr=2\pi\times\unit[124]{Hz}$ to
$\wr=2\pi\times\unit[207]{Hz}$. The line is an exponential fit to
the data. The inset shows the evolution of the mean temperature of
the cloud.} \label{fig:elasticrelax}
\end{figure}

Fig.\,\ref{fig:elasticrelax} shows a typical relaxation
measurement for a cloud with a mean temperature of
$\bar{T}(t)=(2T_r(t)+T_z(t))/3\approx\unit[42]{\uK}$. In the
inset, we have plotted the evolution of the mean temperature where
the slight increase for long thermalization times can be
attributed to dipolar relaxation collisions. An exponential fit to
$\DTts$, where $t^*$ is the rescaled time, yields a relaxation
time constant $\taurel$. We have experimentally verified, that the
relaxation time constant is inversely proportional to the density
of the cloud, thus ruling out anharmonic mixing.

The experimentally observed rethermalization rate is related to
the atomic collision rate by $\Gacoll=\alpha\Garel$, where
$\alpha$ is in general a temperature-dependent proportionality
factor. The elastic collision rate \Gacoll in a non-degenerate
ultra-cold gas in thermal equilibrium is given by
$\Gacoll=\bar{n}(t)\EW{\sigma(v_r)v_r}_\mathrm{th}$, where
$\sigma(v_r)$ is the velocity dependent $s$-wave elastic cross
section, $\EW{\cdot}_\mathrm{th}$ denotes thermal averaging and
$\bar{n}=\int n^2(\vecr)dV/\int n(\vecr)dV$ is the mean density.
The proportionality factor  can take values from  $\alpha = 2.65$
for a energy-independent elastic cross section
\cite{Monroe1993a,Arndt1997a,Kavoulakis2000a} to $\alpha=10.7$ for
an energy-dependent cross section in the unitarity limit
\citep{Arndt1997a, Kavoulakis2000a}. A detailed analysis of our
experiment shows that the situation is intermediate. Therefore, we
use the analytical result \cite{Kavoulakis2000a}, Eqs. (72) and
(88)
\begin{equation}\label{relaxfullth}
    \Garel(T)=\frac{1}{4}\bar{n}\EW{\sigma(v_r)
    v_r}_\mathrm{sth}
\end{equation}
for the relaxation rate to analyze our data, where
$\EW{\cdot}_\mathrm{sth}$ denotes the non-standard thermal average
required for relaxation of thermal anisotropies.

In order to obtain the scattering length, we use the
effective-range approximation to the collisional cross-section
\cite{Joachain:1975},
\begin{equation}
    \label{sigmaaEre}
    \sigma =\frac{8\pi a^2}{k^2a^2+(\frac{1}{2}k^2r_e a-1)^2},
\end{equation}
where $k$ is the relative wave vector, $a$ the scattering length
and $r_e$ the effective range of the potential. The latter can be
derived from the scattering length and the $C_6$ coefficient
\cite{Gao:1998a,Flambaum:1999a}. For chromium, we obtain from the
static polarizability \cite{Cambi:1991a} $C_6 \approx
\unit[(1050\pm200)]{a.u.}$, where
$\unit[1]{a.u.}=\unit[9.57\times10^{-80}]{Jm^6}$. In the
temperature range we explore, the non-resonant contribution of
higher order partial waves to the elastic cross section can be
neglected, since for chromium the threshold temperature for
$d$-waves is on the order of \unit[1.3]{mK} \cite{Julienne:1989a}.

\begin{figure}
\includegraphics[width=0.9\columnwidth]{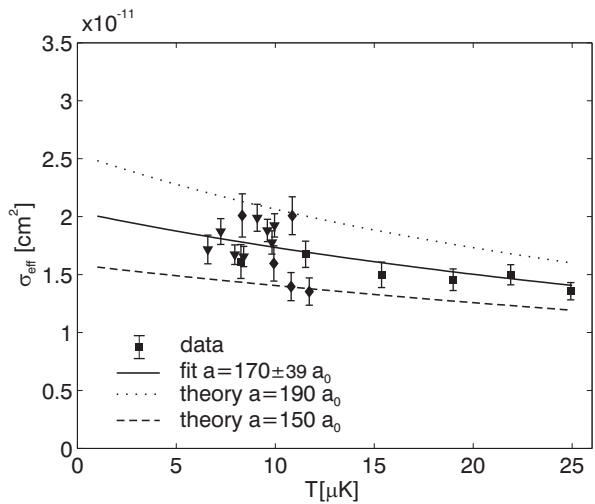}
\caption{Temperature dependence of the effective elastic cross
section of \Crzf for $T\rightarrow 0$. The full line is a fit to
the data yielding $\azf=\unit[(170\pm 39)]{a_0}$. The dotted
(dashed) line is a theoretical curve derived from
Eqs.\,\ref{relaxfullth} and \ref{sigmaeff} for a positive
scattering length of $\unit[190]{a_0}$ ($\unit[150]{a_0}$). The
error bars are derived from statistical and calibration
uncertainties. \label{fig:elasticsigma}}
\end{figure}

The magnitude of the scattering length can be more accurately
determined from low temperature measurements of the elastic cross
section.
In Fig.\,\ref{fig:elasticsigma}, we have plotted the effective
elastic cross section\footnote{The factor $\alpha=2.65$ has been
obtained from \cite{Kavoulakis2000a}, Eq. (88).}
\begin{equation}\label{sigmaeff}
    \sigmaeff(T):=\frac{2.65\,\Garel(T)}{\bar{n}\,\vrth}
\end{equation}
for temperatures below \unit[25]{\uK}. Here $\vrth=\sqrt{{16\kB
T}/{\pi m}}$ is the relative thermal velocity with \kB the
Boltzmann constant and $m$ the chromium mass. The solid curve in
Fig.\,\ref{fig:elasticsigma} is a fit of Eq.\,(\ref{relaxfullth})
to the data points shown, yielding a scattering length of
$\azf=\unit[(170\pm 39)]{a_0}$ corresponding, for the nominal
$C_6$, to an effective range of $r_e=\unit[83]{a_0}$. The given
error combines statistical and systematic uncertainties where the
dominant contribution is from a systematic uncertainty of
\unit[20]{\%} in the determination of the density of the atomic
cloud.
\begin{figure}
\includegraphics[width=0.9\columnwidth]{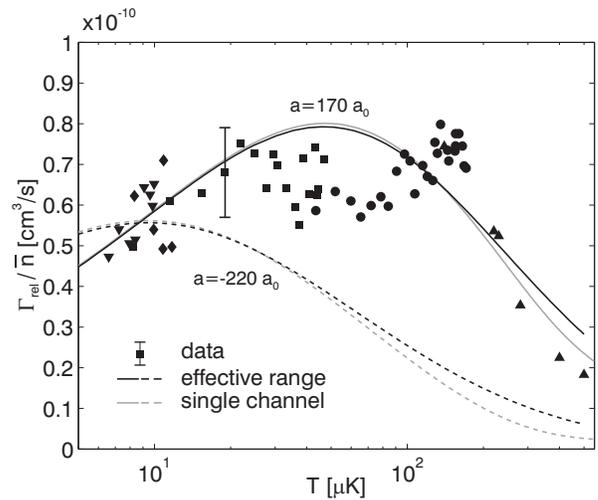}
\caption{Temperature dependence of the density-normalized
rethermalization rate for \Crzf. The black curves are obtained
from Eq. \ref{relaxfullth} and \ref{sigmaaEre}, gray curves are
single-channel calculations. Solid (dashed) curves correspond to a
scattering length of $\unit[170]{a_0}$ ($\unit[-220]{a_0}$). The
data have been compiled from five runs under different
experimental conditions which are represented by different
symbols.} \label{fig:elasticfulllog}
\end{figure}

The sign of the scattering length can be deduced from a comparison
of the measured temperature dependence of the relaxation rate with
theoretical models over a sufficiently large temperature range. We
have performed relaxation rate measurements for ultra-cold \Crzf
covering a temperature range of approximately two orders of
magnitude. The result is shown in Fig.\,\ref{fig:elasticfulllog}
where we have plotted the density-normalized relaxation rate
${\Garel}/{\bar{n}}$ versus the mean temperature of the atomic
sample. We have plotted Eq.\,(\ref{relaxfullth}) for two pairs of
theoretical model cross sections. The black curves correspond to
the effective-range cross section Eq.\,(\ref{sigmaaEre}) and the
gray curves are numerically calculated for a model potential
consisting of a Morse potential well at short range, smoothly
joined to a van der Waals tail at long range. The full curves are
calculations for $\azf=\unit[170]{a_0}$ obtained from the low
temperature fit in Fig.\,\ref{fig:elasticsigma}, whereas we have
assumed a negative scattering length $a=\unit[-220]{a_0}$ and the
corresponding effective range $r_e=\unit[229]{a_0}$ for the dotted
curves. The latter combination of values has been chosen to
achieve the best agreement with the relaxation data at low
temperatures under the condition that $a$ be negative.
Fig.\,\ref{fig:elasticfulllog} shows the main results of the
paper: (i) The comparison between the experimental data and the
theoretical curves clearly shows that the assumption of a negative
scattering length is incompatible with the relaxation data. (ii)
The effective-range cross section agrees very well with the
single-channel calculations over the measured temperature range.
The small deviation of the relaxation rates from the theoretical
curves with $\unit[170]{a_0}$ in the intermediate temperature
range deserves further investigation.
\begin{figure}
\includegraphics[width=0.9\columnwidth]{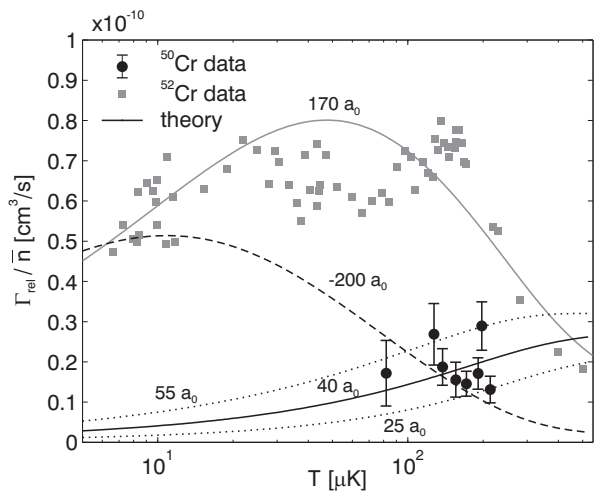} \caption{Comparison
of relaxation rates for \Crf (black circles) and \Crzf (gray
squares) normalized to the density. The black solid curve is a fit
of the single-channel numerical calculation described in the text
to the \Crf data yielding $\af=\unit[(40\pm15)]{a_0}$. All other
curves are the result of single-channel calculations with the
indicated scattering lengths.} \label{fig:elastic5052}
\end{figure}

Knowledge of the scattering lengths for two different isotopes of
the same atomic species can give further insight into the
molecular potentials involved in the collision. We have therefore
performed additional relaxation measurements on \Crf, which is
also a boson with vanishing nuclear spin. In
Fig.\,\ref{fig:elastic5052} we have plotted the results for the
density-normalized relaxation rates for \Crf (black circles). For
comparison, we have included again the data for \Crzf (gray
squares) demonstrating that the relaxation rates for \Crf are much
smaller than those for \Crzf. All curves are obtained from the
numerical single-channel calculation. The black solid curve is a
fit to the \Crf data resulting in a scattering length of
$\af=\unit[(40\pm15)]{a_0}$, where we have again assumed a
systematic uncertainty on the order of \unit[20]{\%}. For the
dashed curve we have assumed a negative scattering length, where
$a=\unit[-200]{a_0}$ gave the best agreement with the data. Mass
scaling our scattering length from \Crzf to \Crf for a variable
number of deca-triplet bound states gives very strong evidence for
a positive sign for \af.

Our analysis yields effective $s$-wave scattering lengths, since
the rather strong magnetic dipole-dipole interactions in chromium
(6 Bohr magnetons) have been neglected. However, preliminary
multi-channel calculations show that the influence of the
spin-dipolar interaction should be within our error bars.

In conclusion, we have measured the temperature-dependence of the
scattering cross section for \Crzf and for \Crf in ultracold
spin-polarized samples. We obtain for the deca-triplet scattering
lengths $\azf=\unit[(170 \pm 39)]{a_0}$ and $\af=\unit[(40 \pm
15)]{a_0}$. The positive sign of \azf was deduced from a simple
effective-range theory. This result together with model
calculations provides strong evidence for a positive sign of \af.
In particular, the reasonably large value of \azf and its sign
make \Crzf a promising candidate for achieving BEC.

 Our findings supplement spectroscopic data on chromium
dimers \cite{Casey:1993a}, which are only available for the
singlet potential and should thus help improve the accuracy of
theoretical calculations of molecular potentials
\citep{Andersson:1995} and of collision properties of chromium.
Improved model potentials based on our results might resolve the
origin of the peak in the inelastic collision rate observed at
\unit[3]{mK} \cite{Carvalho:2003}. Of particular interest would be
the determination of the position and width of magnetically
tunable Feshbach resonances. If Feshbach resonances are readily
accessible, chromium with its large dipole moment would be an
ideal system for studying the interplay between contact and
dipolar interaction since both interactions could then be
experimentally controlled \cite{Giovanazzi:2002a}.
\begin{acknowledgments}
We wish to thank the NIST, Gaithersburg, Atomic Physics Division
led by P. S. Julienne for the ground-state collisions code and F.
Pirani for useful comments on the Cr$_2$ potential. This work was
funded by the Forschergruppe "Quantengase" of the DFG and the
European RTN "Cold Quantum Gases" under Contract No.
HPRN-CT-2000-00125. P.O.S has been supported by the
Studien\-stiftung des deutschen Volkes.
\end{acknowledgments}

\end{document}